\documentclass[11pt]{article}

\usepackage{amssymb}
\usepackage{amsmath}
\usepackage[utf8]{inputenc}
\usepackage{hyperref}
\numberwithin{equation}{section} 

\begin{document}

\title{The Clifford Algebra Approach to Quantum Mechanics B: The Dirac Particle and its relation to the Bohm Approach. \\ }
\author{B. J. Hiley\footnote{E-mail address b.hiley@bbk.ac.uk.}  \; and R. E. Callaghan.}
\date{TPRU, Birkbeck, University of London, Malet Street, \\London WC1E 7HX.}
\maketitle

\begin{abstract}

In this paper we present for the first time a complete description of the Bohm model of the Dirac particle. This result demonstrates again that the common perception that it is not possible to construct a fully relativistic version of the Bohm approach is incorrect.  We obtain the fully relativistic version by using an approach based on Clifford algebras outlined in two earlier papers by Hiley and by Hiley and Callaghan.  The relativistic model is different from the one originally proposed  by Bohm and Hiley and by Doran and Lasenby.  We obtain exact expressions for the Bohm energy-momentum density, a relativistic quantum Hamilton-Jacobi for the conservation of energy which includes an expression for the quantum potential and a relativistic time development equation for the spin vectors of the particle.  We then show that these reduce to the corresponding non-relativistic expressions for the Pauli particle which have already been derived by Bohm, Schiller and Tiomno and in more general form by Hiley and Callaghan. In contrast to the original presentations, there is no need to appeal to classical mechanics at any stage of the development of the formalism.   All the results for the Dirac, Pauli and Schr\"{o}dinger cases are shown to emerge respectively from the hierarchy of Clifford algebras $ {\cal{C}}_{13}, {\cal{C}}_{30}, {\cal{C}}_{01}$ taken over the reals as Hestenes has already argued.  Thus quantum mechanics is emerging from one mathematical structure with no need to appeal to an external Hilbert space with wave functions.

\end{abstract}

\section{Introduction.}

In a recent paper, Hiley and Callaghan \cite{bhbc08} have shown that both the Schr\"{o}dinger and Pauli theories can be completely described within a pair of nested Clifford algebras, ${\cal C}_{0,1}$ and ${\cal C}_{3,0}$ and there is no need to use the wave function.  Thus contrary to popular belief there is no need to regard the Hilbert space as an indispensable basic feature of quantum mechanics. Instead,  the information required to reproduce all quantum effects can be carried within the algebra itself, specifically by appropriate elements of a minimum left ideal as first pointed out by Frescura and Hiley \cite{ffbh80} where the reasons for adopting this approach were discussed.

We showed in our previous paper \cite{bhbc08} that the Clifford algebraic formalism was completely equivalent to the conventional approach to quantum mechanics.  This opens up the possibility of a different interpretation, an interpretation that brings us closer to providing an explanation of quantum phenomena in terms of a non-commutative geometry.  Furthermore, and somewhat surprisingly, we found 
our formulation  enabled us to make contact with the Bohm approach, casting a completely different light on this interpretation.  Specifically in the case of the
Schr\"{o}dinger theory, we obtained exactly the formalism used in the  Bohm model  \cite{db52}, \cite{dbbh93} and \cite{ph93}.  While for a Pauli particle, which of course carries spin, we found a generalisation of the Bohm, Schiller and Tiomno \cite{dbrs1}, \cite{dbrs2} theory.  Both these models are non-relativistic and therefore the previous paper contained no new physics, being of a pedagogical nature and written to show explicitly the role played by the Clifford algebra.  The purpose of this paper is to apply the same methods to the Dirac theory.

In conventional relativistic quantum mechanics, the Clifford algebra made its appearance indirectly as an attempt to remove the negative energy that arises in the relativistic expression for the energy, $E=\pm \sqrt{p^{2} + m^{2}}$. (We use natural units throughout).  Dirac proposed that we start from the  positive linear Hamiltonian $H = \boldsymbol{\alpha.p} + \beta m$ where $\boldsymbol{\alpha}$ and $\beta$ were unknown elements that have to be determined by the constraints placed on the energy.  Thus in order to satisfy  $E^{2}=p^{2}+m^{2}$,  $\boldsymbol{\alpha}$ and $\beta$ must be four anti-commuting objects that could be represented by $4 \times 4$ matrices.  With the linear Hamiltonian, the equation of motion becomes\begin{eqnarray}
i\partial_{t}\Psi = [\boldsymbol{\alpha.p} + \beta m]\Psi				
\end{eqnarray}
where $\Psi$ is a four-dimensional column vector, assumed to be an element in a finite dimensional Hilbert space.  This can be simplified by writing the  equation in manifestly covariant form
\begin{equation}
(i\gamma^{\mu}\partial_{\mu} - m)\Psi=0		\label{eq:dirac}
\end{equation}
This reinforced the belief that the wave function was the natural vehicle for describing quantum phenomena. However what Dirac had actually discovered was that $\boldsymbol{\alpha}$ and $\beta$ were elements of the Clifford algebra  ${\cal C}_{1,3}$.  The Pauli spin matrices had already alerted us to a possible role for the Clifford algebra, but the Schr\"{o}dinger theory seemed not to require a Clifford algebra.  However once one realises that ${\cal{C}}_{0,1}\cong \Bbb C$ then we see that the Schr\"{o}dinger theory can also be discussed in terms of a Clifford algebra.  The way the Clifford structure can be used in the Schr\"{o}dinger and Pauli approaches to quantum theory was described  in detail in a previous paper  \cite{bhbc08}.   The purpose of this paper is to show how these Clifford algebraic techniques can be applied to the Dirac theory.  In order to bring out the quantum aspects of our approach, we will confine our discussion to the free, charge-neutral Dirac particle.  It is easy to extend our approach to charged particles by simply introducing minimal coupling to the electo-magnetic field in the usual way.

\section{Previous Approaches.}


We will assume that the application of the conventional approach through spinors represented in a Hilbert space is by now well known.  However attempts 
to discuss the Dirac theory from within the Clifford algebra itself have been judged to have achieved limited success.  Here we are thinking specifically of the work of Eddington \cite{ae36}, Takabayasi \cite{tt57} and Hestenes \cite{dh03} and references therein.  In this paper we will extend our previous work \cite{bhbc08} to the Dirac theory.  We will also show that, as a consequence, we are able to provide a complete relativistic generalisation of the Bohm approach.

Previous attempts to build a Bohm model have generally been restricted to using the expression for the Dirac current to obtain trajectories for the Dirac electron.   This approach has been used by Bohm and Hiley  \cite{dbbh93}, \cite{dbbh91}  and Holland \cite{ph93}, where some applications have been discussed.   Gull, Lasenby and Doran \cite{sgal93} have also used the Dirac current to calculate trajectories in their investigation of quantum tunnelling in the relativistic domain.  However in no case was a quantum Hamilton-Jacobi equation for the conservation of energy derived and consequently no relativistic expression for the quantum potential has been found.  A very early proposal was made by Bohm himself \cite{db53}, but this proposal was found to be unsatisfactory as will become apparent from the work presented in this paper.
These partial attempts have been seen as a critical weakness of the Bohm model and has led to the general conclusion that the Bohm approach cannot be applied in the relativistic domain.  This paper shows adds further evidence to show that this conclusion is not  correct. 

In view of the arguments proposed in \cite{bhbc08} and in \cite{ffbh80}, we show how we can also dispense with the wave function $\Psi$ in the Dirac theory and instead use an element of a minimal left ideal to carry the information normally carried by the wave function.  Thus once again we can work entirely within the algebra with no need to introduce an external Hilbert space structure.  This, of course, not only provides us with an alternative approach to the Dirac theory itself, but it also provides  a way to generalise the Bohm model so that it can be applied to all relativistic particle situations.   This work complements the work of Horton, Dewdney and Nesteruk \cite{ghcd00} \cite{cdgh06} \cite{cdgh04} who discuss the application of the Bohm approach to relativistic scalar particles and vector bosons.

To obtain our results we must examine the Clifford algebra ${\cal{C}}_{1,3}$ in more detail and abstract all the information necessary to describe a relativistic quantum process.  We should point out that previous attempts at exploiting this Clifford structure have been made by Hestenes \cite{dh}, \cite{dh03}.  The method we use in this paper is more direct, being based on the ideas discussed in Hiley \cite{bh08} and in Hiley and Callaghan \cite{bhbc08}.

\section{The Clifford Algebra ${\cal{C}}_{1,3}$.}


\subsection{${\cal{C}}_{1,3}$ over the reals.}


In order to treat the Dirac theory in the same way as we have treated the Schr\"{o}dinger and Pauli theories, we need first to understand some of the complexities introduced by the larger Clifford algebra ${\cal{C}}_{1,3}$.
This algebra is generated by $ \{1, \gamma_{\mu} \}$
where $\mu=0, 1, 2, 3$ and $[\gamma_{\mu}, \gamma_{\nu}]=2g_{\mu \nu}$.  We choose the diagonal metric $(1, -1, -1, -1)$.  In our approach we are taking the Clifford algebra as basic and projecting out a space-time manifold using the mapping $\eta:\gamma_{\mu}\rightarrow {\hat e}_{\mu}$.  Here ${\hat e}_{\mu}$ is a set of orthonormal unit vectors in a vector space $V_{1,3}$, the Minkowski space-time for an equivalent class of Lorentz observers \cite{bhbc08}.

\subsection{The Primitive Idempotents}

 
 The next step is to examine the primitive idempotent structure in ${\cal{C}}_{1,3}$, as these play a key role in our approach.  In the case of the Schr\"{o}dinger and Pauli algebras, the primitve idempotents have a very simple structure.  For example in the Schr\"{o}dinger algebra, the primitive idempotent is trivial, being the identity of the algebra.  In the Pauli algebra all primitive idempotents can be generated from $\epsilon = (1+\sigma_{i})/2$ where $\sigma_{i}$ is one of the generators of the algebra ${\cal{C}}_{3,0}$. These idempotents form an equivalence class under the transformation $\epsilon' = g\epsilon g^{-1}$ where $g$ is an element of the spin group  $SU(2)$.

In the Clifford algebra ${\cal{C}}_{1,3}$, taken over the reals, we can distinguish three classes of primitive idempotents \cite{ac90}, characterised by
\begin{eqnarray}
\epsilon_{1} =(1+\gamma_{0})/2;\hspace{1cm}\epsilon_{2}=(1+\gamma_{30})/2;\hspace{1cm}\epsilon_{3}=(1+\gamma_{123})/2. \label{eq:setprims}					
\end{eqnarray}
Each of these idempotents can generate an equivalence class using the relations $\epsilon_{i}'=g\epsilon_{i}g^{-1}$ where, in this case, the $g$ are elements of the spin group $SL(2C)$.    Crumeyrolle \cite{ac90} calls the elements in each set  `geometrically equivalent', since they are equivalent under the Lorentz group.  The three defining idempotents can be related to each other through $\epsilon_{i}=u_{ij}\epsilon_{j}u_{ij}^{-1}$ where $u_{ij}(\ne g)$ is some other invertible element of the algebra.

The  reason for choosing one specific idempotent from the  set (\ref{eq:setprims}) is determined by the physics we are interested in. For example,  the choice of $(1+\gamma_{0})/2$ is made in order to pick out a particular Lorentz frame by fixing the time axis defined by $\gamma_{0}$ through the projection $\eta$ defined in a previous section. Any other Lorentz frame can be obtained by choosing the appropriate Lorentz transformation, $g = \Lambda({\bf v})$ from $Sl(2C)$ and forming $\gamma_{0}^{\prime}=\Lambda^{-1}({\bf v})\gamma_{0}\Lambda({\bf v})$.  In this way we relate the algebra to an equivalence class of Lorentz observers.

 Alternatively $(1+\gamma_{30})/2$ can be chosen if we want to pick out  a particular spin direction, the 3-axis in the Lorentz frame defined by $\gamma_{0}$.   In this case  we are highlighting the even sub-algebra ${\cal{C}}_{1,3}^{0}\cong {\cal{C}}_{3,0}$.  Other spin directions can be described by transforming $\gamma_{30}$ using appropriate elements of the spin group.

If we want to consider the little group of the Lorentz group, $SO(2,1)$ \cite{ew59}, then  either $(1+\gamma_{123})/2$ or $(1+\gamma_{30})/2$ can be chosen. These elements project onto the idempotents of the corresponding Clifford algebra ${\cal{C}}_{2,1}$.

As we have remarked above, the three primitive idempotents themselves are equivalent under a larger group, the Clifford group which consists of {\em all} invertible elements of the algebra.  For example we find the following relations
\begin{eqnarray*}
\epsilon_{2}=(1-\gamma_{3})\epsilon_{1}(1+\gamma_{3})/2\\
\epsilon_{3}=(1-\gamma_{5})\epsilon_{1}(1+\gamma_{5})/2 \\
\epsilon_{3}=(1-\gamma_{012})\epsilon_{2}(1+\gamma_{012})/2
\end{eqnarray*}
This group contains reflections and inversions and is thus clearly related to parity and time reversal.  We will not discuss further these relations in this paper.

\subsection{Formation of elements of a minimal left ideal in ${\cal{C}}_{1,3}$ over the Reals.}


We start by choosing the primitive idempotent $\epsilon_{1}=(1+\gamma_{0})/2$.  Thus we are working in a particular Lorentz frame.  A general element of the minimal left ideal is generated by forming ${\cal{A}}(1+\gamma_{0})/2$, $\cal{A}$ being all the elements of ${\cal{C}}_{1,3}$.  This  minimal left element can then be written in the form
\begin{eqnarray}
2\Phi_{L}(x^{\mu})=a(x^{\mu})(1+\gamma_{0}) +b(x^{\mu})(\gamma_{12}+\gamma_{012})+
c(x^{\mu})(\gamma_{23}+\gamma_{023})\nonumber\\
+d(x^{\mu})(\gamma_{13}+\gamma_{013})+f(x^{\mu})(\gamma_{01}-\gamma_{1})+g(x^{\mu})(\gamma_{02}-\gamma_{2})\nonumber\\
+h(x^{\mu})(\gamma_{03}-\gamma_{3})+n(x^{\mu})(\gamma_{5}-\gamma_{123})
\label{eq:minLI}								
\end{eqnarray}
Here the set $\left\{a(x^{\mu}), b(x^{\mu}), c(x^{\mu}), d(x^{\mu}), f(x^{\mu}), g(x^{\mu}), h(x^{\mu}), n(x^{\mu})\right\}$ are the eight real  functions that can be used to specify the quantum state of the Dirac particle.  Recall that this element encodes all the information that is normally encoded in the wave function which uses 4 complex numbers.  Thus there exists a relation between these two sets of parameters which will be given in section 3.7, equation (\ref{eq:apsi}).

Examining (\ref{eq:minLI}) shows that it splits into an even part\footnote{by `even' we mean that $\Phi_{L}^{e}$ only contains even products of the $\gamma s$.}, $\Phi_{L}^{e}$,  and odd part, $\Phi_{L}^{o}$, with the same set of parameters occurring in each expression.  Since the same information about the system is contained in each part, it would be convenient if we could use only one of these expressions in describing the state of the process.  We will follow Hestenes \cite{dh75} and use only  $\Phi_{L}^{e}$ which, for convenience, we will write as $\phi_{L}$. Specifically we write
\begin{eqnarray}
 \phi_{L}=a+ b\gamma_{12}+ c\gamma_{23}+ d\gamma_{13}+ f\gamma_{01}+ g\gamma_{02}+ h\gamma_{03}+ n\gamma_{5}
 \label{eq:dhideal}								
\end{eqnarray}
The set $\{a, b, c, d, f, g, h, n\}$ is  determined by solving the Dirac equation as was first shown by Sauter \cite{s}. 

We can put these relations in a simplified and more transparent form by writing   the basis of this ideal as 
\begin{eqnarray*}
[1, \gamma_{12}, \gamma_{23}, \gamma_{13}, \gamma_{01}, \gamma_{02}, \gamma_{03}, \gamma_{5}]\;(1 + \gamma_{0}).
\end{eqnarray*}
Then we can split this expression into two parts 
\begin{eqnarray}
[1, \gamma_{12}, \gamma_{23}, \gamma_{13}]\;(1+\gamma_{0})
\label{eq:large}									
\end{eqnarray}
and
\begin{eqnarray}
[\gamma_{5},\gamma_{01},\gamma_{02},\gamma_{03}]\;(1+\gamma_{0})=\gamma_{5}[1, \gamma_{12}, \gamma_{23}, \gamma_{13}]\;(1+\gamma_{0})	\label{eq:small}								
\end{eqnarray}
It should be noted that the square bracket in (\ref{eq:large}) and the second square bracket in (\ref{eq:small})  are the same and clearly generate the quarternion algebra.   Thus the elements of the ideal split into two quaternions, explaining the old name `bi-quternion' used by Eddington \cite{ae36}.  Splitting the element  in this fashion immediately enables us to distinguish between its large and small components.  To see this recall that an infinitesimal Lorentz boost, $\delta v$, along, say the $x$-axis, can always be written in the form $\Lambda (\delta v)=(1+\gamma_{01}\delta v)$ so that when $\delta v \rightarrow 0$ the contribution of the $\gamma_{01}$ is negligible.  Thus, in general, when the boost is negligible, the contribution of the $\gamma_{0i}$ terms will be negligible, hence the term `small'.  One we can show that this division corresponds exactly to the conventional division of the Dirac wave function into its large and small components.

\subsection{Real or Complex field?}


There is a difficulty in taking ${\cal{C}}_{1,3}$ over the reals which we must now address.  In the Schr\"{o}dinger and Pauli cases we noted that there was no need to introduce the complex numbers in order to compare with the standard approach.  In both these cases, the Schr\"{o}dinger Clifford, ${\cal{C}}_{0,1}$, and the Pauli Clifford, ${\cal{C}}_{3,0}$, contain an element in the centre of the algebra that squares to minus one.  In the case of ${\cal{C}}_{0,1}$ it was, trivially, the generator of the algebra $e$,  while for the Pauli, ${\cal{C}}_{3,0}$, it was the element $e_{123}$, the product of all three generators.  The Clifford algebra ${\cal{C}}_{1,3}$, only has the identity in the centre, therefore we cannot simply use ${\cal{C}}_{1,3}$ over the reals.  This would make it impossible to reproduce all the results of the standard theory which exploits $\Bbb C \otimes {\cal{C}}_{1,3}$.

However if we go to the next larger Clifford algebra, which is  ${\cal{C}}_{4,1}$.  We find this algebra does contain a non-trivial term in the centre, namely, $F_{01234}\equiv F_{5}$, where $F_{A}\;(A=0,1\dots4)$ are the generators of the algebra ${\cal{C}}_{4,1}$, a notation introduced by Rocha and Vaz \cite{rrjv07}. We will now show how we can still use the real number field and yet account for all the properties of the standard approach.  To do this we will exploit the isomorphism ${\cal{C}}_{4,1}\cong \Bbb C \otimes {\cal{C}}_{1,3}$.  This will enable us to use ${\cal{C}}_{4,1}$ taken over the reals. The structure is now rich enough for us to be able to  relate the eight {\em real} parameters $\{a, \dots ,n\}$ to the four complex numbers, $\psi_{1}, \psi_{2}, \psi_{3}$ and $\psi_{4}$,  contained in the standard Dirac vector in Hilbert space (i.e. the Dirac spinor). 

In ${\cal{C}}_{4,1}$ the generators  $F_{A}$ satisfy the anti-commutaion relations $[F_{A},F_{B}]_{+}=2G_{AB}$ where  $G_{AB}$ is the metric tensor with diagonal elements $(-1, 1, 1, 1, 1)$.
Here the element $F_{5}=F_{01234}$, being in the centre of the algebra, will play the role of the imaginary $i$.  Specifically we use the projection $\rho:{\cal{C}}_{4,1}\rightarrow  \Bbb C \otimes {\cal{C}}_{1,3}$ defined by,
\begin{eqnarray}
\rho(F_{\mu 4})=\gamma_{\mu}\hspace{1cm}\mu = 0, 1, 2, 3. \label{eq:gamu}\\
\rho(F_{5})=i \hspace{3.6cm}       \label{eq:pro}
\end{eqnarray}						
One significant point to notice is the way the signature changes from $(4,1)$ to $(1,3)$.  This is characteristic of the way the next lower Clifford algebra is embedded in the even part of the higher algebra.  This result follows immediately from the nature of the projection $\rho$ as we will now show.

Consider the anti-commutator
\begin{eqnarray*}
[F_{\mu 4}, F_{\nu 4}]_{+}=-2G_{\mu \nu}\hspace{1cm}\mu = 0, 1, 2, 3.
\end{eqnarray*}
Then forming
\begin{eqnarray*}
[\rho(F_{\mu 4}),\rho(F_{\nu 4})]_{+}=-2\rho(G_{\mu \nu})
\end{eqnarray*}
and using equations (\ref{eq:gamu}) and (\ref{eq:pro}) we find
\begin{eqnarray*}
[\gamma_{\mu},\gamma_{\nu}]_{+}=-2G_{\mu \nu}=2g_{\mu \nu}
\end{eqnarray*}
where $g_{\mu \nu}$ is the metric of ${\cal{C}}_{1,3}$.  This shows why the the signs of the metric alternate.

Let us now consider the Pauli algebra ${\cal{C}}_{3,0}$ which is an even sub-algebra of ${\cal{C}}_{1,3}$ under the projection
\begin{eqnarray*}
\lambda(\gamma_{k0})=\sigma_{k} \; \\
\lambda(\gamma_{5})=\sigma_{123}
\end{eqnarray*}
In this case ${\cal{C}}^{0}_{1,3}\cong{\cal{C}}_{3,0}$ giving the metric tensor as $g_{ij}=(1,1,1)$.  Following this through to the lowest algebra ${\cal{C}}_{0,1}$ we have the  sequence of embeddings
\begin{eqnarray*}
{\cal{C}}_{2,4}\rightarrow{\cal{C}}_{4,1}\rightarrow{\cal{C}}_{1,3}\rightarrow{\cal{C}}_{3,0}\rightarrow{\cal{C}}_{0,2}\rightarrow{\cal{C}}_{0,1}.
\end{eqnarray*}
Here we have chosen to start this sequence with the conformal Clifford ${\cal{C}}_{2,4}$ since this contains the Penrose twistors \cite{rp67}.  Physically this means we have a hierarchy of Clifford algebras which fit naturally the physical sequence:

 Twistors $\rightarrow$ relativistic particle with spin $\rightarrow$ non-relativistic particle with spin $\rightarrow$ non-relativistic particle without spin.

\subsection{The Dirac algebra}


Let us now examine the structure of ${\cal{C}}_{4,1}$ in more detail.   Again we find it contains three classes of primitive idempotent characterised by $(1+F_{4})(1+F_{014})/4, (1+F_{04})(1+F_{014})/4$ and $(1+F_{014})(1+F_{1234})/4$.  All other primitive idempotents can be derived from these using elements of the spin group.  Out of all the possible idempotents we choose the specific idempotent
\begin{eqnarray*}
\epsilon_{F}=(1+F_{04})(1-F_{034})/4=(1+F_{04})(1+F_{5}F_{12})/4
\end{eqnarray*}
If we now use the projection $\rho$ defined in \eqref{eq:gamu} and \eqref{eq:pro} we obtain the primitive idempotent that we will be working with, namely\footnote{It looks as if we have introduced the complex number $i$.  However we are actually using $F_{5}$, which for short hand convenience, we are calling $i$.},
\begin{eqnarray}
\epsilon_{\gamma}=(1+\gamma_{0})(1+i\gamma_{12})/4.  \label{eq:iddirac}
\end{eqnarray}								
The reason for our choice now becomes apparent.  The first term in $\epsilon_{\gamma}$ allows us to project out a shadow manifold, which as we have shown, is equivalent to picking  out a particular Lorentz frame by choosing a given time axis.  The second term allows us to pick out a spin plane which is perpendicular to the spin axis $S_{3}$.  This will enable us to see exactly how the Pauli and the Schr\"{o}dinger Clifford sub-algebras are contained in the larger algebra.  This particular choice of idempotent allows us to use directly the relation of the hierarchy of Clifford algebras discussed in the previous sub-section.  Having said this, it is important to realize that, mathematically, we could choose any primitive idempotent, but we have chosen (\ref{eq:iddirac}) because it enables us to compare our results directly with those of the conventional approach using the Dirac representation as we will show in section 3.7.

\subsection{The Minimum Left Ideal Generated by $\epsilon_{\gamma}.$}


At this stage we need to relate the minimal ideal given in \eqref{eq:dhideal} to  the minimum left ideal generated by the idempotent $\epsilon_{\gamma}$ \eqref{eq:iddirac}.  It can be shown by straight forward multiplication that the basis of the minimal left ideal generated  in this way is
\begin{eqnarray*}
\epsilon_{\gamma}&=&(1+\gamma_{0}+i\gamma_{12}+i\gamma_{012})/4;\\
\gamma_{23}\epsilon_{\gamma}&=&(\gamma_{23}+\gamma_{023}+i\gamma_{13}+i\gamma_{013})/4;\\
\gamma_{03}\epsilon_{\gamma}&=& (\gamma_{03}-\gamma_{3}+i\gamma_{5}-i\gamma_{123})/4;\\
\gamma_{01}\epsilon_{\gamma}&=& (\gamma_{01}-\gamma_{1}- i\gamma_{02}+i\gamma_{2})/4.		
\end{eqnarray*}
To make contact with the element of the ideal shown in equation \eqref{eq:minLI}, we write
\begin{eqnarray} 
4\Phi_{L_{\gamma}}=(a-ib)\epsilon_{\gamma}+(c-id)\gamma_{23}\epsilon_{\gamma}+(h-in)\gamma_{30}\epsilon_{\gamma}+(f+ig)\gamma_{01}\epsilon_{\gamma}.
\label{eq:copLI}							
\end{eqnarray}
To check the relation of this element with that given in \eqref{eq:minLI}  we must expand out this expression and examine the even part.  We find immediately that $\Phi_{L}=2\Re[\Phi_{L_{\gamma}}]$. If we examine the imaginary part of $\Phi_{L_{\gamma}}$ we find that  by multiplying from the left by $\gamma_{21}$, we have $\Re[\Psi_{L_{\gamma}}] =\gamma_{21}\Im[\Phi_{L_{\gamma}}]$.  Thus the same information is encoded four times in the minimum left ideal generated by $\epsilon_{\gamma}$ given by \eqref{eq:iddirac}.

\subsection{The Standard Matrix Representation}


Since this paper is mainly about the use of Clifford algebras in physics, we need to relate the information contained in elements of minimal left ideals with that contained in the wave function.  For comparison we choose the standard representation used by Dirac \cite{pd}.  In this representation we have
\begin{eqnarray}
\gamma_{0} =\begin{pmatrix}
   1   &  0  \\
     0 &  -1
\end{pmatrix}
\hspace{0.5cm} \gamma_{j}=\begin{pmatrix}
   0   &  \sigma_{j}  \\
  -\sigma_{j}    &  0
\end{pmatrix}
\hspace{0.5cm}\gamma_{5}=\begin{pmatrix}
   0   &  -i  \\
    -i  &  0
\end{pmatrix}   								
\end{eqnarray}
Here we write $\gamma_{5}=\gamma_{0123}$. 

If we substitute these matrices into the equation \eqref{eq:copLI} and compare the resulting matrix with the usual Hilbert space column vector $\Psi$ with components $ \psi_{1}, \psi_{2},\psi_{3},\psi_{4},$, we find 
\begin{eqnarray}
\psi_{1} = a-ib;\hspace{0.5cm}\psi_{2} = -d-ic,\hspace{0.5cm}\psi_{3}= h-in,\hspace{0.5cm} \psi_{4} = f+ig,  \label{eq:apsi} 		
\end{eqnarray}
This will enable us the write
\begin{eqnarray*}
2a=(\psi_{1}^{*}+\psi_{1});\;\;2ib=(\psi_{1}^{*}-\psi_{1});
\;\;-2d=(\psi_{2}^{*}+\psi_{2});\;\;2ic=(\psi_{2}^{*}-\psi_{2});\\
2h=(\psi_{3}^{*}+\psi_{3});\;\;2in=(\psi_{3}^{*}-\psi_{3});\;\;\;2f = (\psi_{4}^{*}+\psi_{4});\;\;\;2ig = (\psi_{4}-\psi_{4}^{*}).\hspace{0.1cm}
\end{eqnarray*}
We will use these relations later in the paper to compare our results  with those of  the standard approach. 

\subsection{Minimal Right Ideals.}


In an earlier paper \cite{bh08} we argued the once we have decided on a particular idempotent we need to construct the {\em Clifford density element} [CDE], $\rho_{c}=\Phi_{L}\Phi_{R}$, where $\Phi_{R}$ is the conjugate to $\Phi_{L}$.  There are two ways to construct this conjugate.  Firstly we can use the direct method by forming  the product $(1+\gamma_{0}){\cal{A}}$ so that 
\begin{eqnarray}
2\Phi_{R}=a(1+\gamma_{0}) - b(\gamma_{12}+\gamma_{012})-
c(\gamma_{23}- \gamma_{023})- d(\gamma_{13}- \gamma_{013})\hspace{0.7cm}\nonumber\\- f(\gamma_{01}-\gamma_{1}) - g(\gamma_{02}-\gamma_{2})- h(\gamma_{03}-\gamma_{3})+n(\gamma_{5}-\gamma_{123})
\label{eq:minRI}							
\end{eqnarray}
 Again we see this element of the right ideal splits into an even and odd part, containing the same information so we will once again consider only the even part, $\Psi_{R}^{0}$ which we will again write as $\psi_{R}$ for convenience. Then
\begin{eqnarray}
\phi_{R}=a - b\gamma_{12}- c\gamma_{23}- d\gamma_{13}- f\gamma_{01} - g\gamma_{02}- h\gamma_{03}+n\gamma_{5}
\label{eq:minRIeven}	
\end{eqnarray}
We can obtain this result more simply by taking the  Clifford anti-morphism called Clifford conjugation\footnote{Any element, $C$,  of the algebra is, in general, a sum of its scalar, vector, bivector, axial vector and pseudo-scalar parts, viz, $C=S+V+B+A+P$.  The Clifford conjugate is given by $\widetilde C=S-V-B+A+P.$} of $\Psi_{L}$.  Symbolically this is written as  $\Psi_{R} = \widetilde\Psi_{L}$, where  $\sim$ denotes the anti-morphism\footnote{Throughout the rest of this paper we will write $\Psi_{R}$ for $\widetilde\Psi_{L}$.}.

\subsection{The Relation of $\Psi_{R}$ to the Dirac Adjoint Spinor}


Now let us return to the relation between the element of the right ideal and the adjoint spinor as defined in standard Dirac theory. We need to start with the element of the left ideal defined in equation \eqref{eq:copLI} and find the conjugate minimum right ideal which also involves  taking the complex conjugate because $\widetilde{F_{5}}=-F_{5}$ so that $\Phi_{R}={\widetilde\Phi_{L}}^{*}$.  Thus we have
\begin{eqnarray*}
4\Phi_{R} = (a+ib)\phi_{1} - (c+id)\phi_{2} - (h+in)\phi_{3} -(f-ig)\phi_{4}
\end{eqnarray*}
where\begin{eqnarray*}
\phi_{1}&=&(1+\gamma_{0}+i\gamma_{12}+i\gamma_{012});\\
\phi_{2}&=&(\gamma_{23}+ \gamma_{023}- i\gamma_{13}- i\gamma_{013});\\
\phi_{3}&=& (\gamma_{03}+\gamma_{3}+ i\gamma_{5}+ i\gamma_{123});\\
\phi_{4}&=& (\gamma_{01}+\gamma_{1}+ i\gamma_{02}+ i\gamma_{2}).
\end{eqnarray*}
  If we now use the relations \eqref{eq:apsi}, we find
\begin{eqnarray}
4\Phi_{R}=\psi^{*}_{1}E_{1} +\psi^{*}_{2}E_{2} -\psi^{*}_{3}E_{3} -\psi^{*}_{4}E_{4}.				
\end{eqnarray}
where the $\{\phi_{i}\}$ have been replaced by their matrix representation counterparts, $\{E_{i}\}$ which are the row vectors $E_{1}=(1\; 0\; 0\; 0); E_{2}=-i(0 \;1\; 0\; 0); E_{3}=(0\; 0 \;1\; 0)$ and $E_{4}=(0\; 0\; 0\; 1)$.  This then gives the expected bilinear invariant 
\begin{eqnarray*}
\Phi_{R}\Phi_{L} = |\psi_{1}|^{2}+|\psi_{2}|^{2}-|\psi_{3}|^{2}-|\psi_{4}|^{2}
\end{eqnarray*}
In the standard Hilbert space approach this corresponds to 
$\bar{\psi}\psi$ where $\bar{\psi}$ is the adjoint wave function, $\bar\psi=\psi^{\dag}\gamma_{0}$.  Thus we see that we can identify the content of the element of the right ideal that we have constructed with the content of the adjoint Dirac spinor.

\section{The Clifford Density Element and Bilinear Invariants}


\subsection{The Clifford Density Element}


Having defined an element of a minimal left ideal and its conjugate right element, we can now form the key quantity, the  Clifford density element, $\rho_{c} = \Phi_{L}\Phi_{R} = \phi_{L}\epsilon\phi_{R}$.  
From the discussion in the previous section, we can write
\begin{eqnarray*}
\phi_{L}\phi_{R}=(a^{2}+b^{2}+c^{2}+d^{2}-f^{2}-g^{2}-h^{2}-n^{2})+2(an-bh-cf+dg)\gamma_{5}.
\end{eqnarray*}
It is notationally more convenient to write $\phi_{L}$ in the form
\begin{eqnarray}
\phi_{L}=Re^{\gamma_{5}\beta/2}U   \label{eq:hesspinor}  
\end{eqnarray}
where $U$ is a bivector such that $U{\widetilde U} =1$.  It is not difficult to show that $U$ is a general element of the spin group, $SL(2C)$ which is, of course, the covering group of $SO(1,3)$.  

The expression (\ref{eq:hesspinor})  is identical to what  Hestenes \cite{dh03} called the `spinor operator'.  As far as the work we a reporting here, we do not use the term `spinor operator' because the notion of an operator has no meaning in the algebraic structure we are discussing.  For us it is simply the even part of an element of the minimal left ideal $\Phi_{L}= Re^{\gamma_{5}\beta/2}U(1+\gamma_{0})/2$.     

 In this notation, the element of the corresponding right ideal is
\begin{eqnarray}
\phi_{R}={\widetilde U}Re^{\gamma_{5}\beta/2}    			
\end{eqnarray}
so that
\begin{eqnarray}
\rho_{c}=\Phi_{L}\Phi_{R}=\phi_{L}(1+\gamma_{0})\phi_{R} \nonumber\\
=R^{2}e^{\gamma_{5}\beta}(1+U\gamma_{0}\widetilde U)		
\end{eqnarray}
With this notation we find
\begin{eqnarray*}
{\phi_{L}}\phi_{R}=R^{2}\cos\beta +R^{2}\gamma_{5}\sin \beta
\end{eqnarray*}

\subsection{The Bilinear Invariants}

  
In order to make contact with the physical parameters that describe the quantum process, we construct bilinear invariants directly from the CDE, $\rho_{c}$.  We have already seen that we define this element through
\begin{eqnarray*}
\rho_{c}=\Phi_{L}\Phi_{R}=\phi_{L}\epsilon\phi_{R}.
\end{eqnarray*}
Now we have the freedom to choose the idempotent $\epsilon$, but as we pointed out earlier, we want to compare our results with the results obtained from the standard approach so we will choose
\begin{eqnarray}
\epsilon_{\gamma}=(1+\gamma_{0}+i\gamma_{12}+i\gamma_{012})/4  			\label{eq:idempot}		
\end{eqnarray}
The set of 16 bilinear elements written in the conventional theory are
\begin{eqnarray*}
\Omega &=& \langle\bar{\Psi}|\Psi\rangle\\
\widehat{\Omega}&=& \langle\bar{\Psi}|\gamma^{5}|\Psi\rangle \\
J^{\mu}&=&\langle\bar{\Psi}|\gamma^{\mu}|\Psi\rangle\\
2S^{\mu\nu}&=&i\langle\bar{\Psi}|\gamma^{[\mu\nu]}|\Psi\rangle\\
\widehat{J}^{\mu}&=&-\langle\bar{\Psi}|\widehat{\gamma}^{\mu}|\Psi\rangle
\end{eqnarray*}
where $\widehat{\gamma}^{\mu}=i\gamma^{5}\gamma^{\mu}$.

One obvious bilinear invariant we need is the Dirac current, $
J^{\mu}=\langle\bar{\Psi}|\gamma^{\mu}|\Psi\rangle$.
In the algebraic approach this current is given by
\begin{eqnarray*}
J^{\mu}=tr(\gamma^{\mu}\phi_{L}\epsilon\phi_{R})
\end{eqnarray*}
Here for simplicity we write $tr$ for the normal trace divided by four.  The only non-vanishing part of the trace comes from the vector part of $\phi_{L}\epsilon\phi_{R}$, that is from $\phi_{L}\gamma_{0}\phi_{R}$.  Let us write
\begin{eqnarray*}
\phi_{L}\gamma_{0}\phi_{R}=\sum A^{\nu}\gamma_{\nu}.
\end{eqnarray*}
Then
\begin{eqnarray*}
J^{\mu}=\sum A^{\nu}tr(\gamma^{\mu}\gamma_{\nu})=A^{\mu}.
\end{eqnarray*}
A tedious but straight forward calculation shows that
\begin{eqnarray*}
A^{0}&=&a^{2}+b^{2}+c^{2}+d^{2}+f^{2}+g^{2}+h^{2}+n^{2}\\
A^{1}&=&2(af-bg+cn-dg)\\
A^{2}&=&2(ag+bf-ch-dn)\\
A^{3}&=&2(ah+bn+cg+df)
\end{eqnarray*}
Then using the relations between the parameters $\{a, b\dots n\}$ and the components of the standard Dirac wave function $\{ \psi_{1}, \dots \psi_{4}\}$ defined in equation (\ref{eq:apsi}), we can write the Dirac current in the standard representation, viz
\begin{eqnarray*}
J^{0}&=&|\psi_{1}|^{2}+|\psi_{2}|^{2}+|\psi_{3}|^{2}+|\psi_{4}|^{2}\\
J^{1}&=&\psi_{1}\psi_{4}^{*}+\psi_{2}\psi_{3}^{*}+\psi_{3}\psi_{2}^{*}+\psi_{4}\psi_{1}^{*}\\
J^{2}&=&i[\psi_{1}\psi_{4}^{*}-\psi_{2}\psi_{3}^{*}+\psi_{3}\psi_{2}^{*}-\psi_{4}\psi_{1}^{*}]\\
J^{3}&=&\psi_{1}\psi_{3}^{*}-\psi_{2}\psi_{4}^{*}+\psi_{3}\psi_{1}^{*}-\psi_{4}\psi_{2}^{*}
\end{eqnarray*}
Thus we have shown that the Dirac current can be written more simply in the algebraic form
\begin{eqnarray}
J=\phi_{L}\gamma^{0}\phi_{R}.		\label{eq:current}  
\end{eqnarray}
This form for the current has already been used by Hestenes \cite{dh73} and Lounesto \cite{pl97}.

In our approach the Dirac current is calculated by simply using the second term in the idempotent  (\ref{eq:idempot}).  It is then useful to ask what the other terms produce when taken individually.  We find that $\phi_{L}\phi_{R}$ produces a scalar and a pseudo-scalar, viz,
\begin{eqnarray}
\phi_{L}\phi_{R}=\Omega + \widehat{\Omega}\gamma_{5}.	\label{eq:scalar}								
\end{eqnarray}

A vital property for the Dirac particle is its spin.  Following Messiah \cite{am62} we find the total angular momentum is
\begin{eqnarray*}
{\bf J}={\bf L}+{\bf S}\hspace{1cm}\mbox{with}\hspace{1cm} {\bf L} = {\bf r} \times {\bf p},\hspace{0.5cm} {\bf S}={\boldsymbol \sigma}/2
\end{eqnarray*}
from which we define the spin bivector as $2S^{\mu\nu}=i\langle\bar{\Psi}|\gamma^{\mu}\gamma^{\nu}|\Psi\rangle$.  In the algebraic approach this spin bivector can be written in the form 
\begin{eqnarray}
2\rho S=\phi_{L}\gamma_{12}\phi_{R}.	\label{eq:2spin}	
\end{eqnarray}
The spatial components of the spin given by
$s^{i}=S^{jk}$ with $i,j,k$ cyclic.

To complete the set of 16 elements we also have the dual of the current, $\hat{J}^{\mu}$, sometimes referred to as the axial vector current or even the Proca current.  Algebraically this is obtained from the expression
\begin{eqnarray}
\hat{J}=\gamma_{5}\phi_{L}\gamma_{012}\phi_{R}=\phi_{L}\gamma_{3}\phi_{R}
\end{eqnarray}						
We have now formed the 16 bilinear invariants that are defined by the CDE, $\rho_{c}$.  It should be noticed that these invariants are obtained from each term in the idempotent so that
\begin{eqnarray*}
 \Omega + \widehat{\Omega}\gamma^{5}&=&\phi_{L}\phi_{R}.\hspace{5.7cm}\\
J&=&\phi_{L}\gamma_{0}\phi_{R}.\\
2S&=&\phi_{L}\gamma_{12}\phi_{R}\\
\hat{J}&=&\gamma^{5}\psi_{L}\gamma_{012}\phi_{R}=\phi_{L}\gamma_{3}\phi_{R}
\end{eqnarray*}Thus we see all the information about the bilinear invariants is contained in the CDE, $\rho_{c}=\Phi_{L}(1+\gamma_{0}+i\gamma_{12}+i\gamma_{012})\Phi_{R}$, each element in the idempotent giving a bilinear invariant.  This means that in our approach the quantum state of the process is described directly in terms of these invariants and is therefore directly observable unlike the wave function.  To find the tensor components  to match up with the conventional theory, we have
\begin{eqnarray*}
\Omega &=& tr[\phi_{L}\phi_{R}]\\
\widehat{\Omega}&=& tr[ \gamma^{5}\phi_{L}\phi_{R}]\\
J^{\mu}&=& tr[\gamma^{\mu}\phi_{L}\gamma_{0}\phi_{R}]\\
2S^{\mu\nu}&=&i tr[\gamma^{[\mu\nu]}\phi_{L}\gamma_{12}\phi_{R}]\\
\widehat{J}^{\mu}&=&-tr[\gamma^{\sigma\mu\nu}\phi_{L}\gamma_{012}\phi_{R}]\hspace{1cm}(\sigma<\mu<\nu).
\end{eqnarray*}

If we try to specify the quantum state of the process using these 16 invariants, all observable quantities, we will find the system  over specified since we only need  8 real parameters $\{a, b, c, d, f, g, h, n\}$ to define the quantum state.  However, as Takabasayi \cite{tt57} has already pointed out, these bilinear invariants are not linear independent.  In fact he shows that  there are only 7 independent bilinear invariants, finding 9 subsidiary conditions relating the invariants. 
We list these here for convenience.
\begin{eqnarray*}
J^{\mu}J_{\mu}&=&-\rho^{2}\\
\hat{J}^{\mu}\hat{J}_{\mu}&=&\rho^{2}\\
J^{\mu}\hat{J}_{\mu}&=&0\\
\mbox{and}\hspace{1cm}\rho^{2}S_{\mu\nu}&=&-\widehat{\Omega}J_{[\mu}\hat{J}_{\nu]}+i\Omega J_{[\sigma}\hat{J}_{\lambda]},\hspace{1cm}\mu,\nu,\sigma,\lambda \:\mbox{cyclic}
\end{eqnarray*}
where $\rho^{2}=\Omega^{2}+\widehat{\Omega}^{2}$  is a positive definite scalar. A more detailed examination of $\rho$ shows that it corresponds to $\psi^{\dag}\psi$ where $\psi$ is the conventional wave function. Thus we can interpret it as  the probability of finding the particle at a particular position.  For a more detailed discussion of the relations between these invariants we refer the interested reader to Takabayasi's paper \cite{tt57}.

Since we only have seven independent bilinear invariants, something is missing. We need one extra bit of information to completely pin down the quantum state.  Takabayasi suggests that we need to construct a set of {\em bilinear invariants of a second kind}.   These are defined as follows
\begin{eqnarray*}
 \bar{\Psi}\Gamma\overleftrightarrow{\partial_{\mu}}\Psi =\partial_{\mu}{\bar{\Psi}}\Gamma\Psi-{\bar{\Psi}}\Gamma\partial_{\mu}\Psi .
 \end{eqnarray*} 
where $\Gamma$ is one of  the sixteen independent terms formed by products of the generators $\gamma^{\mu}$. Here we are using a short hand notation sometimes used in field theory (see Raymond \cite{pr81}). 

Since we need only one type-two invariant to complete our description of the state, we have a choice.  Our choice, which is different from the one chosen by Takayabasi, is the energy-momentum tensor
\begin{equation}
2iT^{\mu\nu}= {\bar{\psi}}\gamma^{\mu}(\partial_{\nu}\psi)-(\partial_{\nu}{\bar{\psi}})\gamma^{\mu}\psi	\label{eq:tmunu}	
\end{equation}
where  $\psi$ is the conventional wave function spinor, while $\bar{\psi}$, is the adjoint spinor. As we will show in the next section, this will help us identify the relativistic generalisation of the Bohm energy and the Bohm momentum that we have used in the Schr\"{o}dinger and Pauli theories, showing that these parameters are not arbitrary but are needed to complete the specification of the quantum state.

\subsection{The Energy-momentum Density.}


Our first step is to write the energy-momentum tensor \eqref{eq:tmunu}
 in algebraic form.  This then becomes
\begin{eqnarray*}
2iT^{\mu\nu}=tr\left\{\gamma^{\mu}[(\partial^{\nu}\phi_{L})\epsilon\phi_{R}-\phi_{L}\epsilon(\partial^{\nu}\phi_{R})]\right\}=tr[\gamma^{\mu}(\phi_{L}\epsilon\overleftrightarrow{\partial^{\nu}}\phi_{R})]
\end{eqnarray*}
Since the only non-vanishing trace is a Clifford scalar, and since $\gamma^{\mu}$ is a Clifford vector, we must find the Clifford vector part of 
$\phi_{L}\epsilon\overleftrightarrow{\partial^{\nu}}\phi_{R}$. We are again using the idempotent given in equation (\ref{eq:iddirac}), namely
\begin{eqnarray*}
\epsilon'_{\gamma}=(1+\gamma_{0}+i\gamma_{12} +i\gamma_{012})/4. 
\end{eqnarray*}
We find the only term in $\epsilon'_{\gamma}$ that gives a Clifford vector is $\gamma_{012}$, so that we need only consider 
\begin{eqnarray}
2T^{\mu\nu}=tr[\gamma^{\mu}(\phi_{L}\overleftrightarrow{\partial^{\nu}}\gamma_{012}\phi_{R})]. 		\label{eq:Tmunu}														
\end{eqnarray}

Now let us first find the energy density $T^{00}$.  In order to do this we need to evaluate $(\phi_{L}\overleftrightarrow{\partial_{\nu}}\gamma_{012}\phi_{R})$ in terms of the functions $\{a(x^{\mu}),b(x^{\mu}),\dots,n(x^{\mu})\}$.  After some straight forward but tedious calculations we find 
\begin{eqnarray}
\phi_{L}\overleftrightarrow{\partial^{\nu}}\gamma_{012}\phi_{R}=A^{\nu}_{i}(x^{\mu})\gamma_{i}			\label{eq:FullTmunu}		  																				
\end{eqnarray}
where the $A_{i}$ are given by
\begin{eqnarray*}
A^{\nu}_{0}&=&-(a\overleftrightarrow{\partial^{\nu}}b+c\overleftrightarrow{\partial^{\nu}}d+f\overleftrightarrow{\partial^{\nu}}g+h\overleftrightarrow{\partial^{\nu}}n)\\
A^{\nu}_{1}&=&-(a\overleftrightarrow{\partial^{\nu}}g+b\overleftrightarrow{\partial^{\nu}}f+c\overleftrightarrow{\partial^{\nu}}h+d\overleftrightarrow{\partial^{\nu}}n)\\
A^{\nu}_{2}&=&\hspace{0.3cm}(a\overleftrightarrow{\partial^{\nu}}f-b\overleftrightarrow{\partial^{\nu}}g-c\overleftrightarrow{\partial^{\nu}}n+d\overleftrightarrow{\partial^{\nu}}h)\\
A^{\nu}_{3}&=&\hspace{0.3cm}(a\overleftrightarrow{\partial^{\nu}}n-b\overleftrightarrow{\partial^{\nu}}h+c\overleftrightarrow{\partial^{\nu}}f-d\overleftrightarrow{\partial^{\nu}}g)
\end{eqnarray*}
If we  use the relations (\ref{eq:apsi}), we can put $T^{00}$ in a more familiar form in terms of the wave function. Thus
\begin{eqnarray*}
T^{00}=i\sum_{j=1}^{4}(\psi^{*}_{j}\partial^{0}\psi_{j}-\psi_{j}\partial^{0}\psi^{*}_{j})=-\sum R_{j}^{2}\partial_{t}S_{j}.
\end{eqnarray*}
where $\psi_{j}$ are the four complex components of the standard Dirac spinor and we have also written $\psi_{j}=R_{j}\exp iS_{j}$ with $R_{j}$ and $S_{j}$ real functions. 

 We claim this is exactly the relativistic version of $\rho E_{B}$ where $E_{B}$ is the Bohm energy density.  To show this, we find that in the non-relativistic limit $ \sum_{j=1}^{j=4} R_{j}^{2}\partial_{t}S_{j} \rightarrow\sum_{j=1}^{j=2}(R_{j}^{2}\partial_{t}S_{j})$.  The latter is just the expression for $\rho E_{B}$  found in the Pauli case.(See Hiley and Callaghan \cite{bhbc08}.)   Furthermore this reduces to $ E_{B}=-\partial_{t}S$, the well known expression for the Bohm energy for the Schr\"{o}dinger particle.  Thus we identify the Bohm energy through the relation
 \begin{eqnarray*}
\rho E_{B} = T^{00}.
\end{eqnarray*}
  Similarly we can also show that the momentum density can be written in the form
\begin{equation*}
T^{k0}=-i\sum_{j=1}^{3}(\psi^{*}_{j}\partial_{k}\psi_{j}-\psi_{j}\partial_{k}\psi^{*}_{j})=\sum R_{j}^{2}\nabla S_{j}.
\end{equation*}
so that we can write  the Bohm momentum as
\begin{eqnarray*}
\rho P^{k}_{B}=T^{k0}
\end{eqnarray*}
Then it is not difficult to show that this again reduces, in the non-relativistic limit, to the Bohm momentum found in the Pauli case and reduces further, if the spin is suppressed, to the well known Schr\"{o}dinger expression $P_{B}=\nabla S$.  This condition is sometimes known as the guidance condition, but here we have no `waves', only process, so this phrase is inappropriate in this context.

To summarise then we have defined the Bohm energy-momentum vector through the relation
\begin{equation}
2\rho P^{\mu}_{B}=2T^{\mu0}=tr[\gamma^{0}(\phi_{L}\overleftrightarrow{\partial_{\mu}}\gamma_{012}\phi_{R})].			\label{eq:BohmP}
\end{equation}

\subsection{The Appearance of Two Currents in the Relativistic Theory.}


In the earlier attempts to apply the Bohm approach to the Dirac theory, Bohm \cite{db53}, Bohm and Hiley \cite{dbbh91} and Gull, Lasenby and Doran, \cite{sgal93} used the Dirac current to provide a means of calculating particle trajectories. This was done under the assumption that these would be a generalisation of the trajectories in the non-relativistic case calculated from the 
 expression for the Bohm energy-momentum density as was investigated by Dewdney et al. \cite{cdph88}.  
 
 The Dirac current is identified with a four momentum $P^{\mu}$ through
\begin{equation*}
\rho P^{\mu}=m(\bar\psi\gamma^{\mu}\psi) 
\end{equation*} 
or in component form
\begin{eqnarray*}
\rho[E, {\bf P}] =\rho m\left[(\psi^{\dag}\beta \psi), (\psi^{\dag}{\boldsymbol \alpha}\psi)\right] .
\end{eqnarray*}
However to show that this is {\em not} the Bohm energy-momentum defined by $\rho P^{\mu}_{B}=T^{\mu0}$  recall that the Dirac equation (\ref{eq:dirac}) can be written in the form 
\begin{equation*}
\gamma^{\mu}(\displaystyle{\not} p\psi)=m\gamma^{\mu}\psi
\end{equation*}
where $\displaystyle{\not} p =i\gamma^{\mu}\partial_{\mu}$. We can now form
\begin{equation*}
\bar{\psi}\gamma^{\mu}(\displaystyle{\not} \overrightarrow p\psi)=m(\bar{\psi}\gamma^{\mu}\psi)
\end{equation*}
together with the adjoint form
\begin{equation*}
(\bar{\psi}\displaystyle{\not} \overleftarrow p)\gamma^{\mu}\psi=m(\bar{\psi}\gamma^{\mu}\psi)
\end{equation*}
Here the Dirac current, $J^{\mu}=(\bar{\psi}\gamma^{\mu}\psi)$, appears on the RHS of both the last two equations.  Thus subtracting these two equations, we find
\begin{eqnarray*}
2mJ^{\mu}&=&\bar{\psi}\gamma^{\mu}(\displaystyle{\not} \overrightarrow p\psi)-(\bar{\psi}\displaystyle{\not} \overleftarrow p)\gamma^{\mu}\psi\\
&=&\bar{\psi}\gamma^{\mu}\gamma^{\nu}(\partial_{\nu}\psi)-(\partial_{\nu}\bar{\psi})\gamma^{\nu}\gamma^{\mu}\psi.
\end{eqnarray*}
Now let us compare this expression with 
\begin{eqnarray*}
2iT^{\mu 0}=\psi^{\dag}(\partial^{\mu}\psi)-(\partial^{\mu}\psi^{\dag})\psi.
\end{eqnarray*}
This expression is clearly not the same as that given for the current $J^{\mu}$.  However expanding the LHS of the equation for $J^{\mu}$, we obtain the Gordon decomposition
\begin{eqnarray*}
2mJ^{\mu}=i\left[\bar{\psi}(\partial^{\mu}\psi)-(\partial^{\mu}\bar{\psi})\psi\right]-i\sum\partial^{\nu}(\bar{\psi}\sigma^{\mu}_{\nu}\psi)
\end{eqnarray*}
where $2i\sigma^{\mu\nu}=(\gamma^{\nu}\gamma^{\mu}-\gamma^{\mu}\gamma^{\nu})$.  This shows that the current can be decomposed into a convection part, the first two terms, and a spin part.  

If we examine the convection part of the current, although it is closer in form to $T^{\mu 0}$, it is still different.  Thus we have two distinct currents, one is the convection current obtained from $J^{\mu}$ and a Bohm energy-momentum density current derived from $T^{\mu 0}$. Both are conserved.  

This difference should not be surprising because the conserved Dirac current, $J^{\mu}$ is the Noether current obtained from a global gauge transformation, while the energy-momentum density is the Noether current produced by a space-time translation.   This difference does not appear in the case of the Pauli and the Schr\"{o}dinger particles, the two currents turn out  to be the same.  

From the physical point of view it seems, at first sight, that having {\em two} conserved currents is strange.  Contemporary field theory emphasizes and uses the Dirac current $J^{\mu}$, since this is term that couples to the electro-magnetic field.  The energy-momentum density is simply used to show that when we integrate it over a volume element, it leads to the global conservation of energy and momentum. [See for example Schewber \cite{ss64}.]  The local energy-momentum density is given no physical meaning.  The main point of the Bohm model is to give meaning to the energy-momentum density of the individual particle.

Attempts have been made to exploit the difference between these two currents in another context, namely, an extended spinning relativistic object.  Takabayasi \cite{tt57},\cite{tt61} and Bohm and Vigier \cite{dbjpv58} have tried to understand this difference by considering a  relativistic liquid drop.  Here they assume that a collection of particles  are in some form of relative      motion.  This implies we have both a circulation of particles and a circulation of energy.  We would expect these to be different in the relativistic domain.  Thus the liquid drop can be described in terms of the relative motion of two centres, the centre of mass-energy and the centre of particle density.  

In the same way we can think of the Dirac particle as an extended process that must be described, in the first approximation, by a process that has two centres of motion which coincide in the rest frame. In the conventional field approach, integrating the energy-momentum density over space avoids raising questions about any possible internal structure of a Dirac particle.  On present experimental evidence this seems to be justified in the case for leptons, although when we come to baryons, we are forced to introduce inner structure in terms of quarks, but the quarks themselves are again assumed to be point-like structures leaving open any deeper structure.  At one level, our investigation here can be regarded as simply exploring the consequences of this bilocal structure.    Let us now move on to consider in more detail the structure of the energy-momentum density and its time evolution.

\section{The Time Evolution of the Energy-Momentum Density.}


\subsection{Symmetrized Energy Constraint}

In order to investigate time development, we need to relate neighbouring points on the base manifold at neighbouring times.  Normally this would be done by using single Dirac derivative.  However because we are exploiting the structure of a Clifford bundle, there are two types of derivatives.  The reason for needing two derivatives was explained in Hiley \cite{bh08}.  These are written in the form
\begin{eqnarray*}
{\overrightarrow D}=\gamma^{\mu}{\overrightarrow\partial}_{\mu}\hspace{1cm}\mbox{and}\hspace{1cm}{\overleftarrow D}={\overleftarrow\partial}_{\mu}\gamma^{\mu}
\end{eqnarray*}
These derivatives give us two time development equations, one equivalent to the Dirac equation which when expressed in terms of an element of a minimal left ideal is
\begin{equation}
i\gamma^{\mu}\overrightarrow{\partial}_{\mu}\Phi_{L}-m\Phi_{L}=0		\label{eq:diracI}								
\end{equation}
The other is equivalent to the adjoint Dirac equation expressed in terms of an element of the corresponding right ideal,
\begin{equation}
i\Phi_{R}\overleftarrow{\partial}_{\mu}\gamma^{\mu}+m\Phi_{R}=0		\label{eq:diracR}								
\end{equation}
The energy constraint insists that we must have $
(\partial_{\mu}\partial^{\mu}+m^{2})\psi=0 $
satisfied for both $\Phi_{L}$ and $\Phi_{R}$, since the quantum state is described by the CDE $\rho_{c}=\Phi_{L}\Phi_{R}$.   Thus two second order derivatives are involved $(\partial_{\mu}\partial^{\mu}\Phi_{L})\Phi_{R}$ and $\Phi_{L}(\partial_{\mu}\partial^{\mu}\Phi_{R})$.  These are used to produce two equations by taking  their sum and their difference.  The sum gives the energy conservation equation 
\begin{eqnarray}
(\partial_{\mu}\partial^{\mu}\Phi_{L})\Phi_{R}+\Phi_{L}(\partial_{\mu}\partial^{\mu}\Phi_{R})+2m^{2}\Phi_{L}\Phi_{R}=0. 		\label{eq:energy}								
\end{eqnarray}
While the difference produces the following equation
\begin{eqnarray}
\Phi_{L}(\partial_{\mu}\partial^{\mu}\Phi_{R})-(\partial_{\mu}\partial^{\mu}\Phi_{L})\Phi_{R}=0				\label{eq:spineq}   
\end{eqnarray}
which, as we will show below, describes the time evolution of the spin and its components.

\subsection{The Quantum Potential}


Now we will use equation (\ref{eq:energy}) to investigate energy conservation.  To analyse this equation further we need to see where  the Bohm energy-momentum as defined in equation (\ref{eq:BohmP}) fits in as was done for the Pauli particle.  To proceed let us first introduce a more general variable $P^{\mu}$ defined by
\begin{eqnarray}
2\rho P^{\mu}=\left[(\partial^{\mu}\phi_{L})\gamma_{012}\phi_{R}-\phi_{L}\gamma_{012}(\partial^{\mu}\phi_{R})\right]    \label{eq:Ptum}
\end{eqnarray}
Let us also introduce a quantity
\begin{eqnarray*}
2\rho W^{\mu}=-\partial^{\mu}(\phi_{L}\gamma_{012}\phi_{R})
\end{eqnarray*}
Combining these two equations, we obtain
\begin{eqnarray*}
(\partial^{\mu}\phi_{L})\gamma_{012}\phi_{R} &=& \rho[P^{\mu}-W^{\mu}]\\
\mbox{and}\quad-\phi_{L}\gamma_{012}(\partial^{\mu}\phi_{R})&=& \rho[P^{\mu}+W^{\mu}]
\end{eqnarray*}
which can then be written as
\begin{eqnarray}
-\partial^{\mu}\phi_{L}&=&[P^{\mu}-W^{\mu}]\phi_{L}\gamma_{012}
\label{eq:W1}
\end{eqnarray}
and
\begin{eqnarray}							
\partial^{\mu}\phi_{R}&=&\gamma_{012}\phi_{R}[P^{\mu}+W^{\mu}] 
\label{eq:W2}							
\end{eqnarray}
Since we are going to use equation (\ref{eq:energy}), we can use equation (\ref{eq:W1}), to form
\begin{eqnarray*}
-\partial_{\mu}\partial^{\mu}\phi_{L}=\left[\partial_{\mu}P^{\mu}-\partial_{\mu}W^{\mu}\right]\phi_{L}\gamma_{012}+\left[P^{\mu}-W^{\mu}\right]\partial_{\mu}\phi_{L}\gamma_{012}
\end{eqnarray*}
After some algebra and finally multiplying from the right by $\psi_{R}$, this equation can be written in the form
\begin{eqnarray}
-(\partial_{\mu}\partial^{\mu}\phi_{L})\phi_{R}= \rho\left[P_{\mu}P^{\mu}+W_{\mu}W^{\mu}-(P^{\mu}W_{\mu}+W_{\mu}P^{\mu})\right]\nonumber\\
+\left[\partial_{\mu}P^{\mu}-\partial_{\mu}W^{\mu}\right](\phi_{L}\gamma_{012}\phi_{R})
\hspace{1.5cm}  					\label{eq:nablaLeft}  
\end{eqnarray}
This gives us the first term in equation (\ref{eq:energy}).  Now we must consider the second term in this equation. Repeating an analogous set of steps but now using equation (\ref{eq:W2}), we find
\begin{eqnarray}
\phi_{L}(\partial_{\mu}\partial^{\mu}\phi_{R})=\rho\left[P_{\mu}P^{\mu}+W_{\mu}W^{\mu}+(P_{\mu}W^{\mu}+W_{\mu}P^{\mu})\right]\nonumber\\
+(\phi_{L}\gamma_{012}\phi_{R})[\partial_{\mu}P^{\mu}+\partial_{\mu}W^{\mu}]\hspace{1.5cm}		\label{eq:nablaRight}   	
\end{eqnarray}
Substituting both these equations in equation (\ref{eq:energy}), we finally find
\begin{eqnarray}
P^{2}+W^{2}+[J\partial_{\mu}P^{\mu}-\partial_{\mu}P^{\mu}J]+[J\partial_{\mu}W^{\mu}+\partial_{\mu}W^{\mu}J]- m^{2}=0	
\label{eq:TotalDiracEnergy}			
\end{eqnarray}
Here we have used the relation $2\rho J=\psi_{L}\gamma_{012}\psi_{R}$, where $J$ is essentially the axial current.  This term reduces to the spin of the Pauli particle in the non-relativistic limit.   Equation (\ref{eq:TotalDiracEnergy}) can be further simplified by splitting it into its Clifford scalar and pseudoscalar parts. The scalar part is
\begin{eqnarray}
P^{2} + W^{2} +[J\partial_{\mu}W^{\mu}+\partial_{\mu}W^{\mu}J]-m^{2}=0				\label{eq:DiracEnergy}		
\end{eqnarray}
This is to be compared with the energy equation 
\begin{eqnarray*}
p_{\mu}p^{\mu}-m^{2} =0
\end{eqnarray*}
Thus we see that the extra two terms must be related to the quantum potential in some way.  Before we arrive at an exact expression for the quantum potential, we must first note that the momentum, $P^{\mu}$, as defined in equation (\ref{eq:Ptum}) is not yet the Bohm momentum defined in equation (\ref{eq:BohmP}).  Equation (\ref{eq:BohmP}) tells us that $P^{\mu}_{B}$ is the $\gamma^{0}$ coefficient in the expression for $P^{\mu}$.  However it is not difficult to abstract the Bohm momentum from the $P^{2}$ term in equation (\ref{eq:DiracEnergy}).  To do this we need to recall equation (\ref{eq:Ptum}) and use equation (\ref{eq:BohmP}) to find
\begin{eqnarray*}
4\rho^{2}P^{2}= \sum_{i=0}^{3} A_{i\nu}A^{\nu}_{i}
\end{eqnarray*}
Using the definition of $P^{\mu}_{B}$ given in equation (\ref{eq:BohmP}), we find
\begin{eqnarray*}
4\rho^{2}P^{2} = 4\rho^{2}P^{2}_{B} + \sum_{i=1}^{3}  A_{i\nu}A^{\nu}_{i}
\end{eqnarray*}
If we write 
\begin{eqnarray*}
\sum_{i=1}^{3}  A_{i\nu}A^{\nu}_{i}= 4\rho^{2}\Pi^{2}
\end{eqnarray*}
we then find equation (\ref{eq:DiracEnergy}) can be written in the form
\begin{eqnarray*}
P^{2}_{B}  + \Pi^{2}+W^{2} +[J\partial_{\mu}W^{\mu}+\partial_{\mu}W^{\mu}J]-m^{2}=0
\end{eqnarray*}
Then we see that the quantum potential for the Dirac particle is
\begin{eqnarray}
Q_{D}=\Pi^{2} + W^{2}+[J\partial_{\mu}W^{\mu}+\partial_{\mu}W^{\mu}J]			\label{eq:QDirac}		
\end{eqnarray}
In the non-relativistic limit, $\Pi = 0$, and equation (\ref{eq:QDirac}) reduces to the quantum potential for the Pauli particle, \cite{bhbc08},  
\begin{eqnarray}
Q_{P}=W^{2} + [S(\nabla W)+(\nabla W)S]	\label{eq:QPauli}   
\end{eqnarray}
where $2\rho S= \psi_{L}e_{12}\psi_{R}$ is the non-relativistic spin limit of $J$.   $W$ is the non-relativistic limit of $W^{\mu}$.

The pseudoscalar part of equation (\ref{eq:TotalDiracEnergy}) is simply $[J\partial_{\mu}P^{\mu}-\partial_{\mu}P^{\mu}J]=0$.  This puts a constraint on the relation between the spin and the momentum of the particle.  In the non-relativistic limit this term vanishes.

\subsection{The Time development of the Spin}


Let us now turn our attention to equation (\ref{eq:spineq}) and show that it leads to an equation for the time development of the spin of the Dirac particle.  By substituting equations (\ref{eq:nablaLeft}) and (\ref{eq:nablaRight}) into equation (\ref{eq:spineq}) we find 
\begin{eqnarray*}
J\cdot\partial_{\mu}P^{\mu}-P\cdot W +J\wedge\partial_{\mu}W^{\mu}=0
\end{eqnarray*}
where we have written
\begin{eqnarray*}
2J\cdot\partial_{\mu}P^{\mu}&=&J\partial_{\mu}P^{\mu}+\partial_{\mu}P^{\mu}J\\
2P\cdot W&=& PW + WP\\
2J\wedge\partial_{\mu}W^{\mu}&=& J\partial_{\mu}W^{\mu}-\partial_{\mu}W^{\mu}J.
\end{eqnarray*}
All of these terms are Clifford bivectors so that equation (\ref{eq:spineq}) gives just one equation.  We can now simplify this equation since
\begin{eqnarray*}
\rho(P\cdot W)= -\partial^{\mu}\rho(P_{\mu}\cdot J)-\rho(P_{\mu}\cdot\partial^{\mu}J)
\end{eqnarray*}
so that 
\begin{eqnarray*}
\partial_{\mu}(\rho P^{\mu})\cdot J+\rho(P_{\mu}\cdot \partial^{\mu}J) +\rho(J\wedge \partial_{\mu}W^{\mu})=0
\end{eqnarray*}
However since $2\rho P^{\mu}=T^{\mu0}$, the conservation of the energy-momentum tensor implies
\begin{eqnarray*}
\partial_{\mu}(T^{\mu0})=2\partial_{\mu}(\rho P^{\mu})=0
\end{eqnarray*}
so that we have finally
\begin{eqnarray}
P_{\mu}\cdot\partial^{\mu}J + J\wedge \partial_{\mu}W^{\mu}=0
\label{eq:spineqn2}  						
\end{eqnarray}
This equation describes the quantum torque experienced by the spin of the particle in the absence of any external field.  Coupling to an external field is achieved in the usual manner by replacing $\partial^{\mu}$ by $\partial^{\mu}-ieA_{\mu}$.  The equation (\ref{eq:spineqn2}) reduces to the quantum torque equation for the Pauli particle \cite{bhbc08}
\begin{eqnarray*}
(\partial_{t} + \frac{P\cdot \nabla}{m})S=\frac{2}{m}(\nabla W\wedge S)
\end{eqnarray*}
Here $P$ is the three-momentum and $S$ and $W$ have the same meanings as in equation (\ref{eq:QPauli}).

\section{Conclusions}


 In this paper we have presented a complete description of the Bohm model of the Dirac particle for the first time.  This result demonstrates that the common perception that it is not possible to construct a fully relativistic version of the Bohm approach is incorrect.  For simplicity we have considered only a neutral Dirac particle in order to bring out the novel quantum features of this approach.  The extension to the case of a charged particle is however straight forward in principle and will not been presented here.   
 
  In detail, we have obtained expressions for the Bohm energy-momentum density and for the quantum potential for the neutral Dirac particle.  We have shown that these quantities become identical to those found for the Pauli particle first presented in \cite{dbrs1} and recently re-derived in more general terms of the Clifford algebra ${\cal C}_{3,0}$ over the reals in \cite{bhbc08}.  We have also obtained the fully relativistic expression for the time evolution of the spin components of the Dirac particle.  We have shown how this reduces to the  non-relativistic version presented in both the above papers.  The numerical details of this non-relativistic version have been presented in Dewdney et al. \cite{cdph88}.  We will examine numerical details of the relativistic  case in another paper.

It should be noted that the approach to the Dirac particle presented here is different from that presented in Bohm and Hiley \cite{dbbh93} and in Doran and Lasenby \cite{sgal93}, \cite{cdal03}.  In these approaches it is the Dirac current $\phi_{L}\gamma_{0}\phi_{R}$ that is used to calculate  trajectories.

It should also be pointed out that Horton, Dewdney and Nesteruk \cite{ghcd00}, although not motivated by the formulation of the theory in terms of Clifford Algebras, have proposed to base a Bohm-type  theory on the use the eigenvectors and eigenvalues of the energy momentum tensor to define particle trajectories and densities for massive relativistic scalar particle using the Klein-Gordon equation \cite{ghcd00}  and vector bosons \cite{cdgh06}. They also give explicit examples of the different trajectories associated with the current (which yields space-like motions) and with the flows of energy momentum (which are necessarily time-like) for the case of scattering from square potentials. A similar approach has also been employed to develop a covariant version of the Bohm field ontology approach to quantised scalar fields \cite{cdgh04}.

  In this paper we have again emphasised that the Dirac current is different from the Bohm energy-momentum current.  This implies that, in general, it possible to generate two different sets of trajectories.  The appearance of two sets of trajectories is a novel feature of the relativistic domain.  This difference is not unexpected as the earlier work of Takabayasi \cite{tt57} has already anticipated this feature.  So too did Tucker \cite{rt88} who used the K\"{a}hler-Atiyah algebra, a somewhat different approach to Clifford algebras, to show how these trajectories arise from Killing vectors.  What our results show is that the difference disappears in the non-relativistic limit so the earlier work did not anticipate this possibility of doubling.  We discuss the reason for this difference but feel that a more detailed investigation is necessary.  We will leave this discussion to another paper.

\section{Acknowledgments} 

We would like to express our thanks to Ernst Binz, Ray Brummelhuis, Malcolm Coupland, Chris Dewdney, Clive Kilmister, Kris Krough and Graham Yendall for many useful discussions.



\end{document}